\documentclass[
aps, pre,
preprint,
onecolumn,
showpacs, 
nopreprintnumbers, 
amsmath,amssymb,
letteraper
]{revtex4}

\usepackage{graphicx}
\usepackage{dcolumn}
\usepackage{bm}

\begin{document}


\title{
An alternative implementation of the Lanczos algorithm for wave function
propagation
}

\author{Quanlin \surname{Jie}}
\email[E-mail: ]{qljie@whu.edu.cn}
\affiliation{%
Department of Physics, Wuhan University,
Wuhan 430072, P. R. China
}%

\author{Dunhuan \surname{Liu}}
\affiliation{%
Department of Physics, Wuhan University,
Wuhan 430072, P. R. China
}%

\date{\today}

\begin{abstract} 
We reformulate the Lanczos algorithm for quantum wave function propagation in
terms of variational principle. By including some basis states of previous
time steps into the variational subspace, the resultant accuracy increases by
several orders. Numerical errors of the alternative method accumulate much
slower than that of the original Lanczos method.  There is almost no extra
numeric cost for the gaining of the accuracy, i.e., the accuracy increase
needs no extra operations of the Hamiltonian acting on state vectors, which
are the major numeric cost for wave function propagation. A wave packet moving
in a 2-dimensional H\'enon-Heiles model serves as an illustration.  This
method is suitable for small time step propagation of quantum wave functions
in large scale time dependent calculations where the operations of the
Hamiltonian acting on state vectors are expensive.
\end{abstract}

\maketitle

\section{introduction}

Propagation of quantum wave functions, i.e., direct integration of the time
dependent Schr\"odinger equation, is a fundamental numeric task. This time
dependent method exhibits many numeric advantages in first principle
calculations. For example, one is able to extract energy spectrum efficiently
from the correlation function via Fourier transformation~\cite{1}, or more
advanced filter diagonalization algorithm~\cite{2}. The efficiency is more
evident when one needs excited energy spectrum in large scale first principle
calculations.

For wave function propagation, the most expansive numeric operations are
products of the Hamiltonian matrix and state vectors, namely, the Hamiltonian
operator acting on state vectors. It is a long standing efforts to develop
efficient algorithm for wave function propagation that uses minimum number of
such matrix-vector product operations.

Among the popular algorithms, such as split operator method~\cite{1,4,5} and
Chebyshev expansion method~\cite{6}, the Lanczos method~\cite{7,8,9} is a
robust and flexible scheme for wave function propagation~\cite{10}. This
method is in principle applicable to any kind of systems, including the time
dependent Hamiltonian~\cite{11}, and there is virtually no need for preparing
knowledge about the considered system to apply the Lanczos method.
Furthermore, the performance of the Lanczos method is relatively insensitive
to the considered system and the initial wave function. In many cases, the
Lanczos method is the best choice to do time dependent calculations. For
example, if one need to compute a quantity (such as the entropy of a
subsystem) changing continuously with time, and the corresponding Hamiltonian
is not suitable to break into two parts to apply the split operator algorithm,
one may consider to employ the Lanczos method for the task.

The Lanczos algorithm transforms a Hermitian matrix into a tri-diagonal form
iteratively~\cite{12}. It has many applications in first principle
calculations, see, e.g.~\cite{13,14,15,16}.  The basic idea of wave function
propagation by Lanczos method is to solve the Schr\"odinger equation for a
given small time step in a low dimensional subspace, namely, the Krylov
subspace. The basis states of the Krylov subspace, $\{\psi_0,\cdots,\psi_m\}$,
are generated by the Lanczos iteration,
$H\psi_j=\beta_{j-1}\psi_{j-1}+\alpha_i\psi_i+\beta_i\psi_{i+1}$, where
$\alpha_i=\langle \psi_i | H |\psi_i \rangle$ is the expectation value of the
Hamiltonian $H$ with respect to the vector $|\psi_i\rangle$, $\beta_i$ is the
norm of the vector $H\psi_i - \beta_{i-1} \psi_{i-1}-\alpha_i\psi_i$ with
$\beta_0=0$ and $\psi_0$ being the wave function obtained from previous step.

The dimension of the Krylov subspace is usually less than $10$ in most cases.
This dimension depends on the time step and the accuracy requirement. Higher
accuracy needs either small time step or large Krylov subspace.  For a given
dimension of the Krylov subspace, the error accumulates linearly with time in
Lanczos method~\cite{10}. If one needs wave function in longer time scale, one
must increase accuracy of each time step to keep the error of the final wave
function within required range. This means the numeric operations are not
simply linearly proportional to the time. There is an optimal choice for the
dimension of the Krylov subspace and the time step to reach the accuracy
requirement of the final state. However, practical situations, e.g.
calculations of correlation function, often need other time steps.

The dimension of the Krylov subspace, or the number of matrix-vector product
operations in a time step, is a key factor to affect the numeric cost for the
Lanczos propagation scheme. Like other algorithms, the operations of the
Hamiltonian acting on the state vectors are the major numerical cost of Lanczos
propagation scheme.  Higher accuracy demands for more such Hamiltonian
operations. In the view point of efficiency, one should keep the number of the
Hamiltonian operations as small as possible for a given time step and accuracy
requirement.

In this paper, we present an alternative to the Lanczos method. It can improve
the accuracy by several orders without extra matrix-vector product operations.
To this end, we reformulate the Lanczos method in terms of variational
principle with the Krylov subspace being the variational subspace.  The basic
idea of improvement is to enlarge the variational subspace by including basis
states of previous time steps into the variational subspace.  Since the
required matrix elements are already calculated in previous time steps, such
enlargement of the variational subspace has virtually no extra numeric cost.
In fact, including basis states of previous steps into the variational
subspace is an efficient method for iteratively diagonalizing large
matrix~\cite{16a,16b,16c}.

\section{Variational approach to the Lanczos propagation scheme}

We first note that one is able to formulate the Lanczos propagation scheme
from the variational principle. For short time $\Delta t$, one can approximate
the evolution operator $U(t)=\exp(-iH\Delta t/\hbar)$ by a polynomial of the
Hamiltonian operator $H$. In other words, one can approximate the wave
function at time $t+\Delta t$, $\psi(t+\Delta t)=U(\Delta t)\psi(t)$, by a
vector in the Krylov subspace spanned by $\{H^i\psi(t),\ i=0,1,\cdots,n-1\}$,
\begin{equation}\label{eq1}
\psi(t+\Delta t)=\sum_{i=0}^{n-1} c_i H^i\psi(t).
\end{equation}
The expansion coefficients $c_i$ are yet to be determined by the variational
principle,
\begin{equation}
\frac{\delta}{\delta c_i}\langle \psi(t+\Delta t) | i\hbar 
\frac{\partial}{\partial t} - H |\psi(t+\Delta t) \rangle = 0.
\end{equation}
One solves the resultant equation of motion to obtain the expansion
coefficients. Instead of $\{H^i\psi(t)\}$, the basis states in original
Lanczos scheme are generated by the Lanczos iteration, and the resultant
Hamiltonian matrix in the variational subspace is tri-diagonal. Of course, the
subspace generated by Lanczos iteration is the same as that spanned by states
$\{H^i\psi(t)\}$.


Another way of arriving at the ansatz (\ref{eq1}) is to relate $H^i\psi(t)$
with the $i$-th order derivatives of state $\psi(t)$ with respects to the
time, $i\hbar\frac{\partial^i}{\partial t^i}\psi(t)=H^i\psi(t)$. One can
approximate the state at time $t+\Delta t$, $\psi(t+\Delta t)$, by a linear
combination of the state $\psi(t)$ and its derivatives up to order $n$. Such
observation may be useful for time dependent systems.

A direct way to improve accuracy of the Lanczos propagation scheme is to
enlarge the variational subspace, i.e., adding more states into the basis
states of variational subspace. This usually means more numerical operations
of matrix-vector product to obtain the basis states and the correspondent
matrix elements of the Hamiltonian.  However, there exist a way to enlarge the
variational subspace with virtually no extra numeric cost.

We achieve this by making using of matrix-vector product operations of
previous steps. To this end, we first note that one can propagate a wave
function backward from $\psi(t)$ to $\psi(t-\Delta t)$. In other words, a wave
function at time $t-\Delta t$, $\psi(t-\Delta t)$, is approximately a linear
combination of states $\{H^i\psi(t),\ i=0,\cdots,m\}$. Here $m$ is the number
of matrix-vector product operations in a time steps. At time $t$, we already
have states $H^k\psi(t-\Delta t)$, $k=0,1,\cdots,m$. Each of these states is
approximately equivalent to a linear combination of states $\{H^{k+i}\psi(t)$,
$i=0,1,\cdots,m\}$.  If one includes some of these states into the variational
subspace, one has effectively products of higher order power of the
Hamiltonian acting on the state vector $\psi(t)$.  Using the same arguments,
one can also include $H^k\psi(t-2\Delta t)$, $H^k\psi(t-3\Delta t)$, $\cdots$,
into the variational subspace.

Implementation of the above scheme is straightforward. It involves similar
procedures as that of original Lanczos propagation scheme. Each step of
propagation is to solve the Schr\"odinger equation in the variational
subspace. The variational subspace is spanned by the basis states of the
original Krylov subspace, $H^i\psi(t)$, and some basis states of previous time
steps, $H^i\psi(t-k\Delta t)$.  The solution of the Schr\"odinger equation in
the variational subspace determines the expansion coefficients of next step's
wave function with respect to the basis states of the variational subspace.
The practical calculation includes the following steps:

(1) Choose the time step $\Delta t$ and the dimension of the variational
subspace, $n$, as well as the number of matrix-vector product operations, $m$,
for each step. The basis states of the variational subspace are
$\phi^{(i)}_k=H^i\psi(t-k\Delta t)$ with $i=0,1,\cdots,m$, $k=0,1,\cdots, K$.
Here $m<n$, and $\psi(t)$ is the current state at time $t$. For practical
applications, it is enough to set $K=1$ and $n<2m$, i.e., one usually needs
only some of last step's basis states to form the variational subspace. In our
implementation of choosing previous time steps' basis states, the order  is
$H^m\psi(t-\Delta t)$, $H^{m-1} \psi(t-\Delta t)$, $\cdots$. This is because
that $H^m\psi(t-\Delta t)$ is closer to $H^i\psi(t)$, $(i>m)$, than other
states, and has less overlap with the basis states of the original Krylov
subspace, $H^i\psi(t)$, $(i\le m)$.


(2) Calculate the matrix-vector products $\phi^{(i)}_0=H^i\psi(t)$,
$i=1,\cdots,m+1$. These states and some states obtained in previous time steps
form the basis states, $\phi^{(i)}_k=H^i\psi(t-k\Delta t)$, of the variational
subspace. Here $\phi^{(m+1)}_0$ is only for calculation of the Hamiltonian's
matrix elements in the variational subspace.  Calculation of the $m+1$
matrix-vector products in this step consumes the major CPU time of the whole
procedure.

(3) Calculate the matrix elements of the Hamiltonian in the variational
subspace, $\mathcal{H}_{ik,jl}=\langle \Phi^{(i)}_k|H|\Phi^{(j)}_l\rangle
=\langle \Phi^{(i)}_k|\phi^{(j+1)}_l \rangle$, and the overlap between the
basis states, $\mathcal{S}_{ik,jl}=\langle \Phi^{(i)}_k|\Phi^{(j)}_l \rangle$.
Here $\Phi^{(i)}_k$ is normalized form of $\phi^{(i)}_k$. For indices $k>0$,
$l>0$, the matrix elements of $\mathcal{H}$ and $\mathcal S$ are already
calculated in previous time steps, one needs only to calculate the terms
$\mathcal{H}_{i0,j0}$, and $\mathcal{S}_{i0,j0}$ in this step. 
The trade off of reusing previous steps' matrix element is that the basis
states is not orthogonal with each other.

(4) Solve the Schr\"odinger equation in the variational subspace
\begin{equation}\label{eq3}
i\hbar\mathcal{S} \frac{d}{dt}C=\mathcal{H}C
\end{equation}
to obtain the expansion coefficients $C=(c^{(0)}_0,\cdots, c^{(i)}_k,\cdots)^T$
of next time step's wave function with respect to the basis states
$\Phi^{(i)}_k$. The computation cost in this step is negligibly small in
comparison with other step's operations. 

(5) Perform linear combination of the basis states to form next step's wave 
function
\begin{equation}
\psi(t+\Delta t)=\sum_{i,k} c^{(i)}_k \Phi^{(i)}_k.
\end{equation}

At the first time step, $t=0$, there is no previous basis states. All the basis
states are formed by states $H^i\psi_0$, $i=0,1,\cdots,n-1$, with $\psi_0$
being the initial state. In next step, we remove the first $m+1$ states
$\psi_0,H\psi_0,\cdots,H^{m}\psi_0$ from the basis states, and add another
$m+1$ states $\psi(\Delta t) ,H\psi(\Delta t),\cdots,H^{m}\psi(\Delta t)$ into
the basis states. In following time steps, we update the basis states of the
variational subspace in the same way, i.e., replacing $m+1$ oldest basis states
with states $\{H^i\psi(t),\ i=0,\cdots,m\}$.

In case $n=m+1$, i.e., the variational subspace including no previous time
step's  basis states, the above procedure is essentially the same as the
original Lanczos propagation scheme.  In such case, numerical cost, storage
requirement, and resulted accuracy are indeed the same as the original one.
In the original Lanczos scheme, the Hamiltonian matrix in the variational
subspace is tri-diagonal, and the overlap matrix is unit. Since the dimension
of the variational subspace is usually small, such difference results in
virtually no extra numerical cost.  Similar to the original Lanczos scheme,
the above procedure is more suitable for small time step propagation which
needs only small variational subspace.
 
The storage requirement is also similar to the original Lanczos scheme. One
needs to store the basis states of the variational subspace, as well as
information about the Hamiltonian. Other informations, such as matrix elements
of the Hamiltonian in the variational subspace and the overlap matrix, need
little memory.

For an given time step and accuracy requirement, there is an optimal choice of
the dimension, $n$, of the variational subspace and the number, $m$, of the
matrix-vector product operations in a time step.  If $m$ is inadequately
small, i.e., one includes too many previous time steps' basis states into the
variational subspace, the overlap matrix may become singular. This means that
there is a limit accuracy for a given time step $\Delta t$ and the number of
matrix-vector product operations in a time step.  We use this property of the
overlap matrix to determine the dimension $n$ for a given $m$ and $\Delta t$.


\section{Numerical results}

We test the performance of the alternative Lanczos method via H\'enon-Heiles
model. It is a particle of unit mass moving in the 2-dimensional
H\'enon-Heiles potential~\cite{17}, $v(x,y)=\frac12(\omega_x^2 x^2+\omega_y^2
y^2)+\lambda y(x^2+\eta y^2)$, where $\omega_x=1.3$, $\omega_y=0.7$,
$\lambda=-0.1$, $\eta=0.1$, and the Planck constant is set to $1$. This system
has a chaotic classical limit. It is widely used to study the
quantum-classical correspondence. Similar to Ref.~\cite{10}, we estimate the
accuracy of the current method by the overlap between the numeric result and
the ``exact'' result. We obtain the ``exact'' result via Chebyshev expansion
method~\cite{6}. The Chebyshev method is a global propagator that can reach an
accuracy of machine's limit with a single time step.

In Figure 1, we show the auto-correlation function $\langle \psi(t) | \psi(0)
\rangle$, i.e., overlap between initial state $\psi(0)$ and the state at time
$t$, $\psi(t)$.  The initial state is a Gaussian wave packet whose center
positions are $(2.0,2.0)$, and center momentums vanish. We use a $64\times64$
grid to represent the 2-dimensional wave function in spatial representation.
The action of momentum operator on the state is performed via Fast Fourier
Transformation (FFT) to transform the state into momentum representation. Thus
the action of the Hamiltonian operator on a state vector needs two FFTs to
transform the state back and forth between coordinate and momentum
representations.  Such matrix-vector product operation is the major numerical
cost of the wave function propagation.  The thin solid line in Fig. 1 is a well
converged result for comparison.  The dashed line is result of alternative
Lanczos method, and the thick solid line is result of the original Lanczos
method. Both the original and alternative Lanczos methods use 2 matrix-vector
product operations in one time step to obtain the auto-correlation function in
Fig. 1. The time step is $\Delta t=0.02$. It is evident that 2 matrix-vector
product operations are not enough to converge for the original Lanczos method.
In fact, one needs at least 4 to 5 matrix-vector product operations in a time
step to make the original Lanczos method converge. On the other hand, the
dashed line from alternative Lanczos method is almost indistinguishable from
the well converged result. Here we include 8 previous basis states,
$\{H^i\psi(t-k\Delta t)$, $i=0,1,2$, $k=1,2,3\}$, into the variational
subspace ($\psi(t-3\Delta t)$ is not included). The total dimension of the
variational subspace is 11.

By reducing the time step to $\Delta t=0.01$, one can obtain the above
converged auto-correlation with only a single matrix-vector product operation
in a time step. For such time step, we achieve similar accuracy as that in
Fig. 1 by including 5 previous time steps' basis states into the variational
subspace.  In contrast, for such time step and accuracy, the original
Lanczos method still needs about 3 to 4 matrix-vector products in one time
step.  

Generally, one can increase the accuracy with virtually no extra numeric cost
by including more previous basis states into the variational subspace.
However, in calculation of Fig. 1, including more than $8$ previous basis
states makes the overlap matrix singular, i.e., the basis states are no longer
independent from each other. In fact, for a given time step $\Delta t$ and the
number $m$ of matrix-vector product in a time step, there is always a limit
number of previous basis states that one can include into the variational
subspace.  In other words, the time step $\Delta t$ and the number $m$
determine limit of the accuracy.

In our test calculations of Fig. 1, the overlap matrix becomes singular
occasionally. when this happens, one can simply remove the non-independent
basis vectors. This can be done by, e.g., Cholesky decomposition of the
overlap matrix. In our implementations, we replace each non-independent vector
by one more state $H^i\psi(t)$ of the Krylov subspace.  Such treatment
preserves the accuracy at the expense of one extra matrix-vector product.

Practical implementation of the alternative Lanczos method is indeed more
stable and robust than the calculation of Fig. 1.  In fact, calculation of
Fig. 1 includes several previous time steps' basis states into the variational
subspace. This treatment almost reaches the accuracy limit with 2
matrix-vector product operations in a time step. Practically, one needs only
including some of last step's basis states into the variational subspace. This
can increase the accuracy by several orders with virtually no extra numeric
cost. The overlap matrix, and thus the whole numeric procedures, are usually
well behaved.

Fig. 2 shows the numeric errors accumulate with time for situations similar to
practical calculations. The solid lines are results of the alternative
Lanczos method, and the dashed lines are results of corresponding original
Lanczos method which uses the same number of matrix-vector product operations.
Same as Ref.~\cite{10}, we use the overlap between numeric result
$\psi_{numeric}$ and ``exact'' result  $\psi_{exact}$ as the measure of
error, 
\begin{equation} 
E_{rr}=1-\langle\psi_{exact}|\psi_{numeric}\rangle.
\end{equation} 
We obtain the ``exact'' state vector by Chebyshev expansion method with
accuracy of machine's limit. We propagate the ``exact'' states with a time
step $\Delta T=4$, and using 1024 Chebyshev polynomials to expand the
evolution operator $\exp(-iH\Delta T/\hbar)$. This is well beyond the accuracy
requirement of the machine's limit. From our tests, 512 Chebyshev polynomials
are indeed well converged. From top to bottom of fig. 2, $N$ is the total
dimension of the variational subspace for the alternative Lanczos method, and
$m$ is the number of matrix-vector product operations in a time step for both
methods. The time step is set to $\Delta t=0.02$ for both methods.  The
initial states, as well as states representation are the same as that of Fig.
1.  It is easy to see that the behavior of the original Lanczos method is
similar to that described in Ref.~\cite{10}, i.e., the error accumulates about
linearly with the time $t$, and the accuracy increases quickly with the number
$m$.

It is evident that, when including some of last step's basis states into the
variational subspace, the accuracy improves drastically. We see that, for
$m=5$, the alternative method is about $5$ orders more accurate than the
original one after time $t\approx 10^3$. And for $m=6$, the alternative method
is about 4 orders more accurate after time $t\approx 5\times 10^3$. The
original method is well converged within the time scale $t<10^4$ for $m=7$.
Even so, the alternative method is still about one order more accurate after
time $t\approx 2\times 10^4$. Another encouraging property of the alternative
method is that its error accumulates much slower than the original one. This
means numeric result of the alternative method is more reliable in long time
scale.


From Fig. 2 and other test calculations, we conclude that the alternative
Lanczos method is suitable for small step wave function propagation. It
improves the accuracy by several orders with almost no extra numeric cost. In
calculation of Fig. 2, the variational subspace is about 10 dimensional, and
includes only 4 last step's basis states. For such setting, the resultant
overlap matrix and thus the overall numeric procedure are well behaved.  In
fact, the included basis states, $H^i\psi(t-\Delta t)$, from previous step
play the role of High order power of the Hamiltonian acting on the state
vector, $H^i\psi(t)$, of the original Lanczos method. From the
dashed lines in Fig. 2, we see that basis states, $H^i\psi(t)$
($i=0,1,\cdots,m$), of the original Lanczos method span the major part
($>99\%$) of the exact wave function $\psi(t+\Delta t)$. Other basis states,
$H^i\psi(t-\Delta t)$, included from previous step span only very small
portion ($<1\%$) of the exact wave function.  Thus, these included basis
states $H^i\psi(t-\Delta t)$ have a relatively lower accuracy requirement to
represent the high order terms $H^i\psi(t)$, ($i>m$). This explains the success
of the alternative method.

In general, implementation of the alternative method is stable and robust,
provided that the basis states $\{H^i\psi(t)\}$ span major part of the next
step's wave function $\psi(t+\Delta t)$.  In fact, for a given time step
$\Delta t$, the accuracy of both alternative and original Lanczos methods is
determined by the dimension of the variational subspace $n$. One chooses
$\Delta t$ and $n$ in the same way as that of the original Lanczos
algorithm. In the alternative implementation, one must specify additionally
the number, $m$, of matrix-vector production in a time step.  It is usually
enough to set $m$ larger than half of $n$, i.e., $n/2<m<n$.  If $m$ is too
small, and one includes too many basis states, $\{H^i\psi(t-k \Delta t)\}$, of
previous steps into the variational subspace, these basis states may be not
independent. Even this happens, the alternative method still works. If a basis
state from previous step is linearly dependent on other basis states of the
variational subspace, we replace this state by an extra state $H^i\psi(t)$
with $i>m$. This keeps the dimension of the variational subspace, and hence
the accuracy of the implementation. The expense of such treatment is one more
matrix-vector product operation for each linearly dependent state. We
implement this treatment during the Cholesky decomposition of the matrix
$\mathcal{S}$ which is a necessary step to solve Eq. (\ref{eq3}). When the
basis states are linearly dependent, the overlap matrix $\mathcal{S}$ becomes
singular. The Cholesky decomposition of $\mathcal{S}$ can find all the
non-independent states. By properly specify the threshold value of the
Cholesky decomposition, this numerically cheap operation can even find the
states that are close to linear superposition of other basis states.


\section{conclusions}

In summary, we present an alternative to the Lanczos method for quantum wave
function propagation in terms of variational principle. This method
approximates short time evolution operator, $U(t)=\exp(-iH\Delta t/\hbar)$, by
a polynomial of the Hamiltonian. In other words, the wave function
$\psi(t+\Delta t)$, resulted from a small time step propagation from
$\psi(t)$, $\psi(t+\Delta t)=U(\Delta t)\psi(t)$, is approximately a vector in
the Krylov subspace spanned by $\{H^i\psi(t),i=0,\cdots,n-1\}$.  One can
employ the variational principle to determine the expansion coefficients. The
original Lanczos method needs to calculate all the basis states $H^i\psi(t)$
explicitly. Construction of theses basis states is the major numeric cost. The
alternative method needs only to calculate some of the basis states,
$H^i\psi(t)$, $i=0,\cdots,m$, which span the major part of the wave function
$\psi(t+\Delta t)$. We use basis states of previous step to play the role of
the other basis states, $H^i\psi(t)$, $i>m$. Practically, it is enough to
include some of last step's basis states, $H^i\psi(t-\Delta t)$, $i\le m$,
into the variational subspace. The accuracy of the alternative method is
several orders higher than the original Lanczos method with same matrix-vector
product operations in a time step. 

This alternative method is especially efficient for small time step wave
function propagation. The error accumulation in the alternative method is much
slower than that in the original Lanczos method, which increases about linearly
with time. The efficiency of the alternative method comes from the fact that
the basis states included from previous steps only span very small portion of
the wave function, and thus the accuracy requirement for construction of these
basis sates is relatively lower. This alternative method has useful
applications in large scale time dependent calculations in which the numeric
cost for the Hamiltonian acting on state vectors is expensive.

\begin{figure}[h]
\includegraphics[angle=-90,width=\columnwidth,clip]{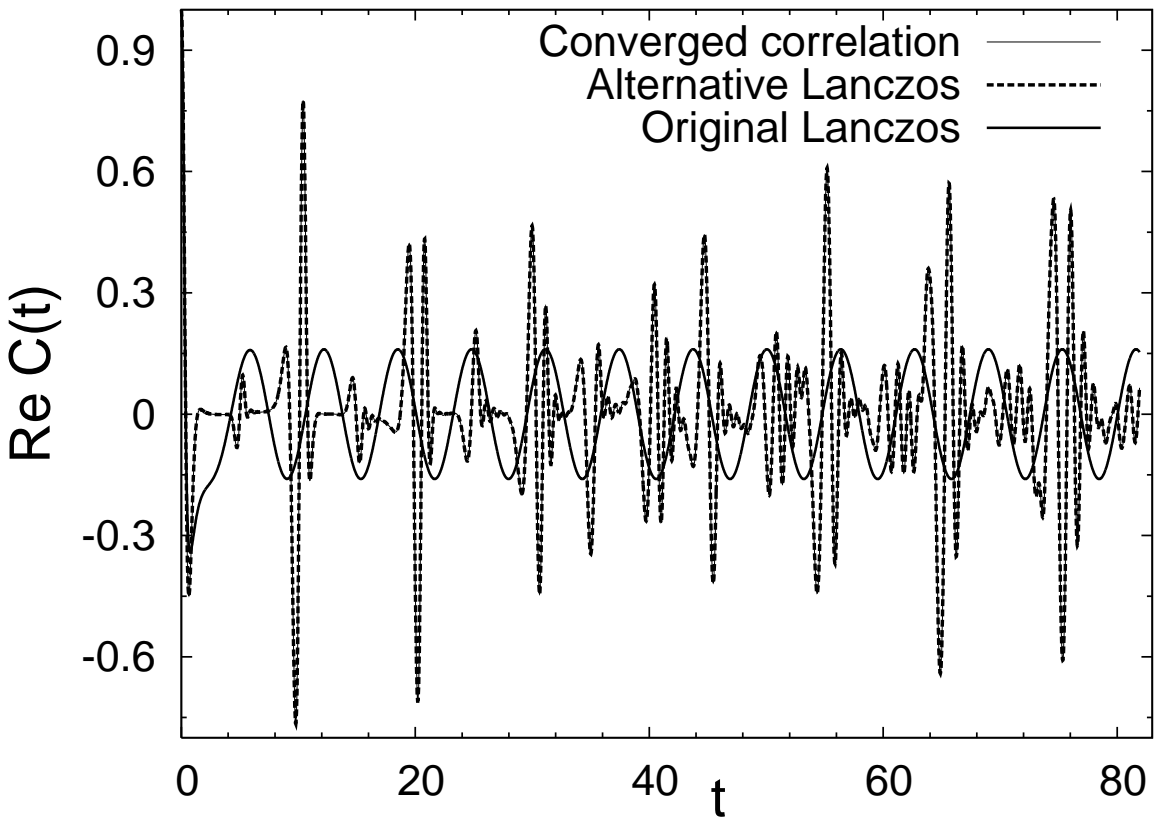}
\label{fig1}
\caption{
Real part of auto-correlation function changes with the time. The
thin solid line is a well converged result. The dashed and thick solid lines
are results of the alternative and original Lanczos methods respectively. Each
time step uses 2 matrix-vector product operations for both dashed and thick
solid lines.
}
\end{figure}

\begin{figure}[h]
\includegraphics[angle=-90,width=\columnwidth,clip]{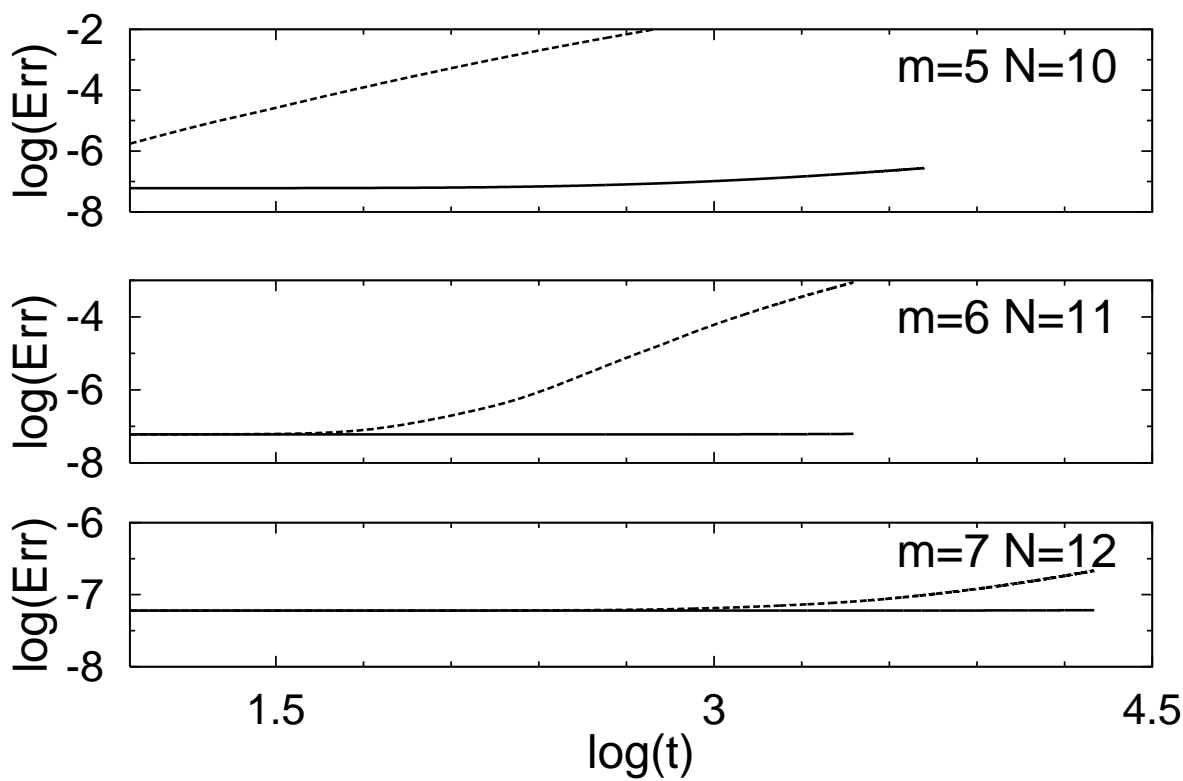}
\label{fig2}
\caption{
Logarithms, base 10, of numeric errors, $Err$, accumulate with the logarithmic
time $t$. Solid and dashed lines are the alternative and original Lanczos
methods, respectively. $m$ denotes the number of matrix-vector product
operations in a time step for both methods, and $N$ is the dimension of the
variational subspace for the alternative method.
}
\end{figure}


\bigskip

This work is supported in part by the National Natural Science Foundation
(Grant No. 10375042), the Research Fund of the State Education Ministry of
China, and the Research Fund of the Wuhan University.


\end{document}